\documentclass[
 reprint,
 amsmath,amssymb,
 aps,
]{revtex4-1}

\usepackage{graphicx}
\usepackage{dcolumn}
\usepackage{hyperref}
\usepackage[normalem]{ulem}
\usepackage{xcolor}
\usepackage{mathrsfs}

\hypersetup{
    unicode=true,           
    pdftoolbar=false,        
    pdfmenubar=false,        
    pdffitwindow=false,      
    pdfstartview={FitH},     
    colorlinks=true,         
    citecolor=blue,           
    linkcolor=blue,
    urlcolor = blue
}

\usepackage[T1,T2A]{fontenc}
\usepackage[utf8]{inputenc}
\usepackage[english]{babel}

\usepackage[capitalize]{cleveref}
\usepackage{etoolbox} 
\usepackage{lipsum} 
\makeatletter
\appto{\appendix}{%
  \@ifstar{\def\thesection{\unskip}\def\theequation@prefix{A.}}%
          {\def\thesection{\Alph {section}}}%
}
\makeatother

\begin{document}

\title{Generation and routing of nanoscale droplet solitons \\ without compensation of magnetic damping}

\author{Andrei I. Nikitchenko}
\author{Nikolay A. Pertsev$^*$}
\affiliation{Ioffe Institute, 194021 St. Petersburg, Russia\\*pertsev.domain@mail.ioffe.ru}

\begin{abstract}
    Magnetic droplet soliton is a localized dynamic spin state which can serve as a nanoscale information carrier and nonlinear oscillator. The present opinion is that the formation of droplet solitons requires the compensation of magnetic damping by a torque created by a spin-polarized electric current or pure spin current. Here we demonstrate theoretically that nanoscale droplet solitons can be generated and routed in ferromagnetic nanostructures with voltage-controlled magnetic anisotropy in the presence of uncompensated magnetic damping. Performing micromagnetic simulations for the MgO/Fe/MgO trilayer with almost  perpendicular-to-plane magnetization, we reveal the formation of the droplet soliton under a nanoscale gate electrode subjected to a sub-nanosecond voltage pulse. The soliton lives up to 50 ns at room temperature and can propagate over micrometer distances in a ferromagnetic waveguide due to nonzero gradient of the demagnetizing field. Furthermore, we show that an electrical routing of the soliton to different outputs of a spintronic device can be realized with the aid of an additional semiconducting nanostripe electrode creating controllable gradient of the perpendicular magnetic anisotropy.
\end{abstract}

\maketitle

\setlength{\parindent}{15pt}
\setlength{\parskip}{0pt}

\section{Introduction}
Magnetic droplet soliton is a type of strongly self-localized mode of magnetization oscillations. Early theoretical studies predicted that a conservative version of such a soliton could form in an ideal ferromagnet having no magnetic damping~\cite{Kosevich1990}. Later, dissipative droplet solitons were described theoretically~\cite{Hoefer2010, Hoefer2012} and demonstrated experimentally in spin transfer nanocontact oscillators (STNOs)~\cite{Mohseni2014, Macia2014} and nanoconstriction-based spin Hall devices~\cite{Divinskiy2018, Dvornik2018} with perpendicular magnetic anisotropy (PMA). In both nanostructures, the formation of solitons is due to the compensation of magnetic damping, which is provided either by the spin-transfer torque generated by a spin-polarized electric current~\cite{Mohseni2014, Macia2014} or by the spin-orbit torque created by a pure spin current injected into the ferromagnet by an adjacent heavy metal~\cite{Divinskiy2018, Dvornik2018}. 

In STNOs, the soliton forms under the nanocontact owing to local magnetization reversal induced by the spin-transfer torque. The droplet size is governed by the nanocontact size and can exceed the latter significantly~\cite{Chung2018} due to the current-induced Zhang-Li torque acting on the droplet boundary~\cite{Li2004, Zhang2004}. The dynamic nature of droplet solitons manifests itself in a large-angle magnetization precession at the droplet boundary. The precession frequency is well below the FMR frequency and practically independent of the driving spin-polarized current~\cite{Hoefer2010, Mohseni2014}. However, it can be tuned by an applied electric field in STNOs with the free layer possessing electric-field-dependent PMA created by an adjacent dielectric layer~\cite{Zheng2020}. 

In this paper, we show theoretically that droplet solitons can be created and routed in ferromagnetic nanolayers without compensation of magnetic damping. This opportunity appears in ferromagnet-dielectric heterostructures having strong interfacial PMA, which can be reduced significantly by an electric field created in the dielectric nanolayer. Such a voltage-controlled magnetic anisotropy (VCMA) represents an efficient tool for the excitation of magnetic dynamics in ferromagnetic nanolayers, including the precessional magnetization switching~\cite{Shiota2012, Kanai2012}, spin reorientation transition (SRT)~\cite{Shiota2009}, coherent magnetization precession~\cite{Nozaki2012, Zhu2012, Viaud2014}, and spin waves~\cite{Verba2014, Rana2017, Nikitchenko2021}. Performing micromagnetic simulations for a perpendicularly magnetized MgO/Fe/MgO trilayer subjected to a weak in-plane magnetic field, we reveal the formation of a droplet soliton induced by a sub-nanosecond voltage pulse locally applied to the MgO nanolayer via a gate electrode. The soliton forms under the gate electrode and lives up to 100~ns at low temperatures, experiencing size oscillations with the period of 0.1-0.5~ns. Furthermore, we demonstrate the propagation of the generated nanoscale soliton over micrometer distances from the nucleation region, which is achieved in a ferromagnetic waveguide owing to the demagnetizing field accelerating the soliton. Finally, an electrical routing of the droplet soliton is realized in a nanostructure with a controllable PMA gradient created by a semiconducting nanostripe electrode. Since only electric fields are needed to generate and route droplet solitons, the proposed technique is distinguished by a low energy consumption, which is advantageous for device applications. 

\section{Results and discussion}
\subsection{Micromagnetic modeling}
\begin{figure}
    \center{
    \includegraphics[height=5cm]{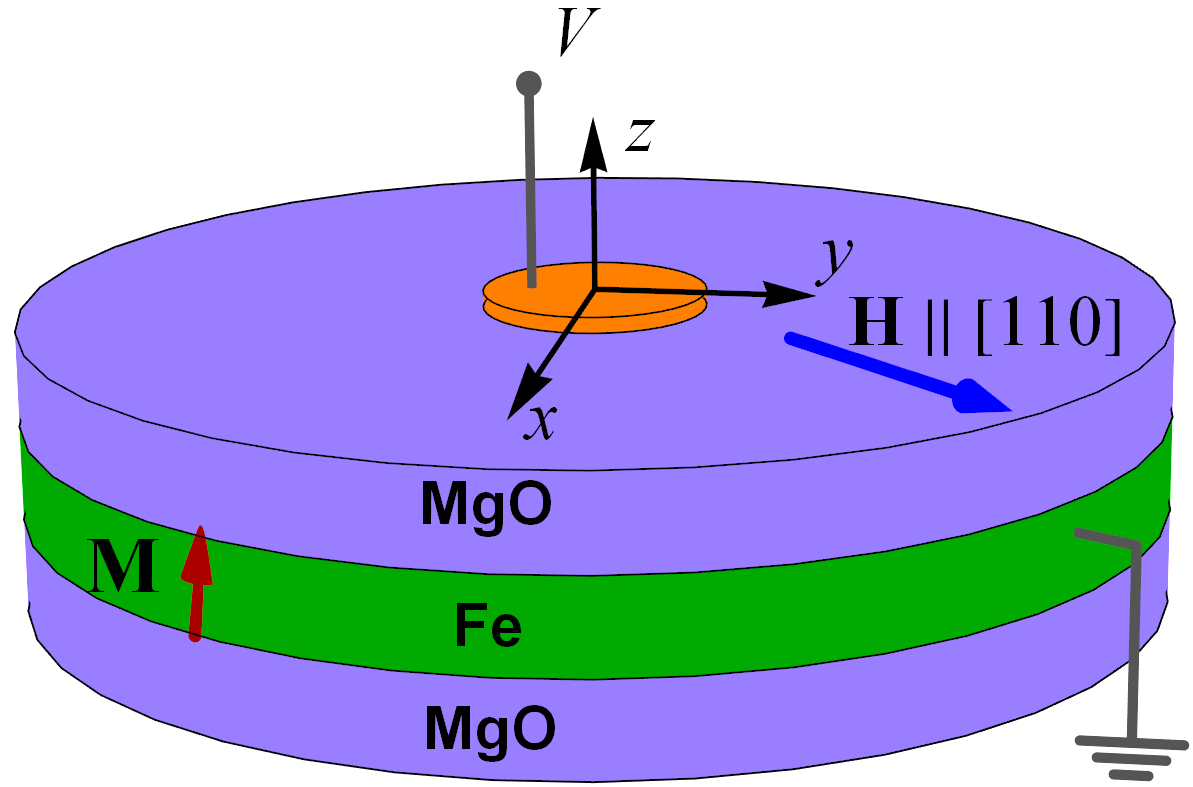}}
    \caption{\label{fig:struct:generation} Cylindrical MgO/Fe/MgO trilayer with a circular gate electrode connected to a voltage source. The Fe nanolayer with a perpendicular magnetic anisotropy is subjected to a weak in-plane magnetic field $\mathbf{H}$, which deflects the magnetization $\mathbf{M}$ from the perpendicular-to-plane orientation. }
\end{figure}

We model the dynamics of the magnetization $\mathbf{M}(\mathbf{r}, t)$ in a (001)-oriented Fe nanolayer grown on MgO(001) and capped with an ultrathin MgO overlayer (Fig.~\ref{fig:struct:generation}). 
\noindent
The chosen Fe thickness $t_\mathrm{F} = 0.75$~nm is smaller than the critical thickness $t_\mathrm{SRT} \approx 0.9$~nm (see Appendix for the calculated critical thickness), below which the perpendicular-to-plane (PP) orientation of $\mathbf{M}$ becomes energetically favorable in the MgO/Fe/MgO structure~\cite{Koziol2013}. The ferromagnetic layer is modeled by a two-dimensional ensemble of $N$ nanoscale computational cells with the sizes $l_x = l_y = 1.5$~nm and $l_z = t_\mathrm{F}$ smaller than the exchange length $\lambda_\mathrm{ex} \approx 3.3$~nm of Fe~\cite{Vaz2008}. Regarding the saturation magnetization $M_s$ as a constant quantity at a given temperature, we calculate the temporal evolution of $\mathbf{M}(\mathbf{r}, t)$ by numerically solving a system of the Landau-Lifshitz-Gilbert (LLG) equations for the unit vectors $\mathbf{m}(\mathbf{r}_n, t) = \mathbf{M}(\mathbf{r}_n, t) / M_s$ defining the magnetization directions in the computational cells situated at the points $\mathbf{r}_n$ ($n=$ 1, 2, 3, ..., $N$). The effective field $\mathbf{H}_\mathrm{eff}$ involved in the LLG equation is written as $\mathbf{H}_\mathrm{eff} = \mathbf{H} + \mathbf{H}_\mathrm{ex} + \mathbf{H}_\mathrm{dip} + \mathbf{H}_\mathrm{an}$, where $\mathbf{H}$ is the external magnetic field, $\mathbf{H}_\mathrm{ex}$ and $\mathbf{H}_\mathrm{dip}$ are the contributions of the exchange and dipolar interactions between spins in Fe, and $\mathbf{H}_\mathrm{an}$ allows for the magnetocrystalline,  magnetoelastic, and interfacial anisotropies existing in the MgO/Fe/MgO structure. The exchange and dipolar contributions to the effective field $\mathbf{H}_\mathrm{eff}$ are calculated as described in our preceding paper~\cite{Nikitchenko2021}, and we use the relation $\mathbf{H}_\mathrm{an} = -(\mu_0 M_s)^{-1} \partial F_\mathrm{an} / \partial \mathbf{m}$ to determine the anisotropy field $\mathbf{H}_\mathrm{an}$ ($\mu_0$ is the magnetic permeability of free space). The effective volumetric energy density $F_\mathrm{an}$ of the magnetic anisotropy can be approximated as~\cite{Pertsev2015} 
\begin{widetext}
\begin{equation}
    \begin{gathered}
        F_\mathrm{an} \approx K_1 (m_x^2m_y^2+m_x^2m_z^2+m_y^2m_z^2) + K_2 m_x^2m_y^2m_z^2  + B_1(u_{xx}m_x^2+u_{yy}m_y^2) \\
        -B_1 \bigg[ \displaystyle\frac{B_1}{6 c_{11}} + \displaystyle\frac{c_{12}}{c_{11}} (u_{xx}+u_{yy}) \bigg]m_z^2 
         + \displaystyle \frac{K_{s\parallel}+K_{s\parallel}'}{t_\mathrm{F}}m_x^2m_y^2 + \displaystyle \frac{K_{s\perp}+K_{s\perp}'}{t_\mathrm{F}}(m_x^2+m_y^2)m_z^2 + \displaystyle\frac{K_s+K_s'}{t_\mathrm{F}}m_z^2,
    \end{gathered}
    \label{eq:Fan}
\end{equation}
\end{widetext}
\noindent
where $K_1$ and $K_2$ are the coefficients of the fourth- and sixth-order terms defining the cubic magnetocrystalline anisotropy of bulk Fe at constant lattice strains $\mathbf{u}$, $B_1$ is the magnetoelastic constant, $c_{11}$ and $c_{12}$ denote the elastic stiffnesses at fixed magnetization, $u_{xx}$ and $u_{yy}$ are the substrate-induced in-plane (IP) strains of the Fe nanolayer, while $K_s$, $K_{s\parallel}$, $K_{s\perp}$ and $K_s'$, $K_{s\parallel}'$, $K_{s\perp}'$ are the parameters characterizing the magnetic anisotropy created by the bottom and top Fe$|$MgO interfaces, respectively. The factor $1/t_\mathrm{F}$ in the last three terms reflects the introduction of only one computational cell in the thickness direction $z$, which is justified by the condition $t_\mathrm{F} < \lambda_\mathrm{ex}/2$.

In our simulations, the numerical integration of the LLG equation is carried out using the projective Runge-Kutta algorithm with the time step of 10~fs, which is much smaller than the duration $\tau_V > 0.1$~ns of the rectangular voltage pulses applied to the gate electrode. To make possible a nonparametric excitation of the magnetic dynamics by VCMA, we introduce an IP magnetic field $\mathbf{H}||[110]$ creating an oblique orientation of the equilibrium magnetization~\cite{Viaud2014}.

\subsection{Electrical generation of magnetic solitons}
We first consider the heterostructure of a circular shape, which includes a nanoscale gate electrode on top of the MgO overlayer (Fig.~\ref{fig:struct:generation}). The electric field $E_z$ created in MgO by a voltage $V(t)$ applied to the gate electrode changes the specific energy associated with the top Fe$|$MgO interface~\cite{Maruyama2009, Niranjan2010}. Therefore, the coefficient $K_s'$ in Eq.~(\ref{eq:Fan}) should be regarded as a voltage-dependent quantity for the computational cells beneath the gate. As $K_{s\perp}'$ and $K_{s\parallel}'$ are much smaller than  $K_s'$~\cite{Pertsev2015}, possible voltage dependences of these parameters can be ignored. Since the dependence $K_s'(E_z)$ is practically linear at the field intensities up to about 2~V~nm$^{-1}$~\cite{Niranjan2010}, the voltage dependence of $K_s'$  can be written as $K_s' = K_s^0 + k_s V/t_\mathrm{MgO}$, where $K_s^0$ is the value of the anisotropy parameter $K_s'$ at $E_z = 0$, $k_s$ is the electric-field sensitivity of $K_s'$~\cite{Pertsev2013}, and $t_\mathrm{MgO}$ is the thickness of the MgO overlayer.

In accordance with the available experimental data~\cite{Koziol2013}, the saturation magnetization of the 0.75-nm-thick Fe film is taken to be $M_s = 1.71\times 10^6$~A~m$^{-1}$. The lattice strains induced in the Fe layer by a thick MgO substrate are set equal to $u_{xx} = u_{yy} = 3.9$\%~\cite{Pertsev2015}. In the numerical calculations, we also use the exchange constant $A_\mathrm{ex} = 20$~pJ~m$^{-1}$~\cite{Vaz2008}, Gilbert damping parameter $\alpha = 0.0025$~\cite{Kamiya2021}, anisotropy coefficients $K_1 = 48$~kJ~m$^{-3}$~\cite{Vaz2008} and $K_2 = 15$~kJ~m$^{-3}$~\cite{Stearns}, PMA parameters $K_s^0 = -9\times 10^{-4}$~J~m$^{-2}$~\cite{Koziol2013} and $K_{s\parallel} = K_{s\perp} = K_{s\parallel}' = K_{s\perp}' = -4.5 \times 10^{-5}$~J~m$^{-2}$~\cite{Pertsev2015}, VCMA coefficient $k_s = 100$~fJ~V$^{-1}$~m$^{-1}$~\cite{Nozaki2016}, magnetoelastic constant $B_1 = -3.3 \times 10^6$~J~m$^{-3}$~\cite{Stearns}, and elastic stiffnesses $c_{11} =  2.42 \times 10^{11}$~N~m$^{-2}$ and $c_{12} =  1.465 \times 10^{11}$~N~m$^{-2}$~\cite{Hirth}. The MgO thickness is set equal to the value of 2~nm, at which the influence of the voltage-induced tunnel current through MgO on the magnetization dynamics can be neglected. The diameter of the MgO/Fe/MgO trilayer is taken to be 450~nm, while the radius $R_G$ of the gate electrode varies from 30 to 120~nm. 

The simulations show that the initial magnetic state of the considered 0.75-nm-thick Fe disk is practically homogeneous. Owing to strong PMA of such a nanolayer, the deviation of the magnetization vector $\mathbf{M}$ from the PP orientation, which is induced by the external IP magnetic field $\mathbf{H}$, appears to be small even at the strongest field $H = 400$~Oe used in our simulations. At the chosen field orientation along the $[110]$ crystallographic direction, which represents the easy axis of the nanolayer's IP anisotropy due to the condition $K_1 + (K_{s\parallel}+K_{s\parallel}') / t_\mathrm{F} < 0$, the mean value of the magnetization polar angle $\theta$ is found to be less than 8$^\circ$ (see Appendix). 

\begin{figure}
    \center{
    \includegraphics[width=1\linewidth]{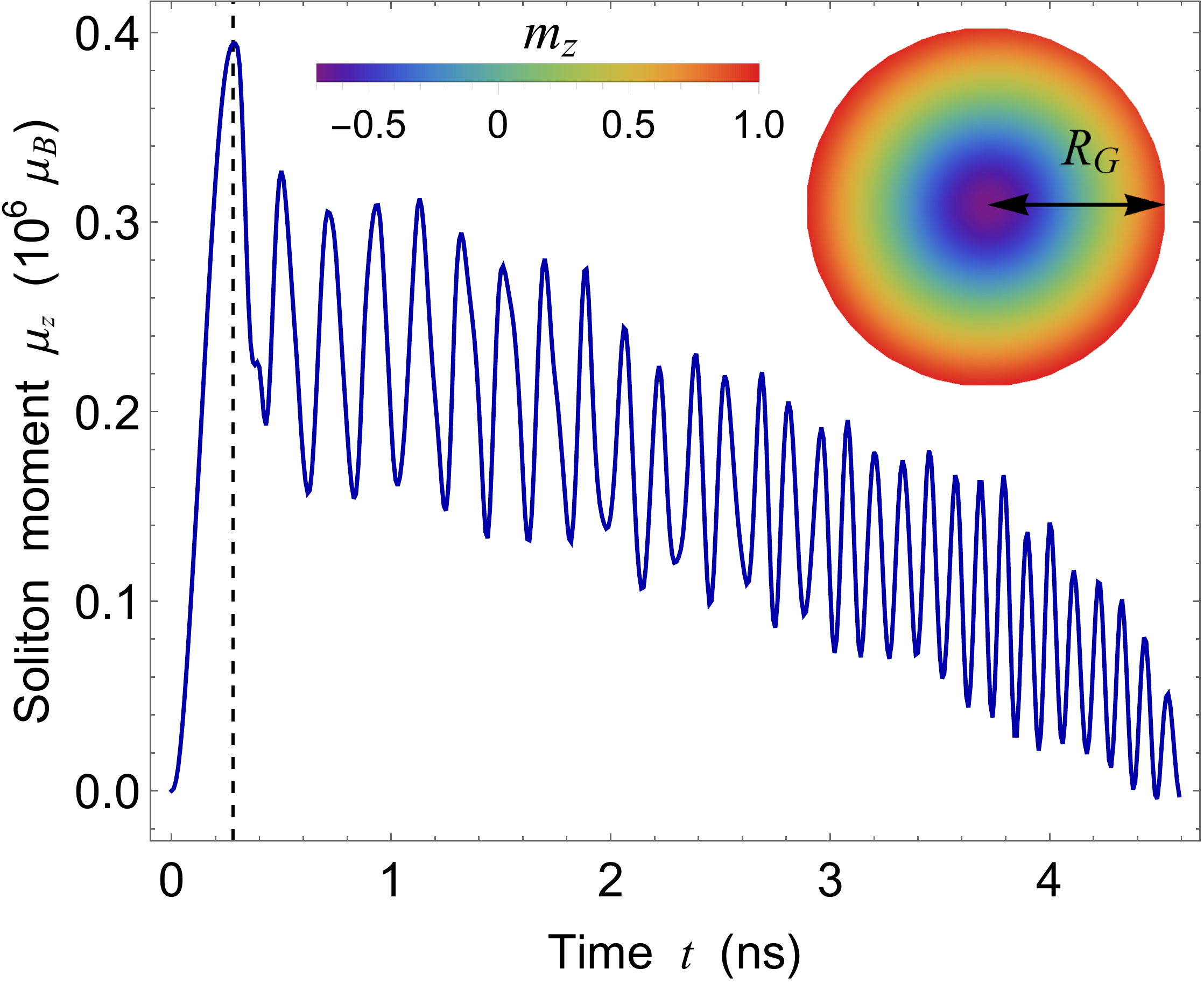}}
    \caption{\label{fig:mu-t-300Oe-50nm} Time dependence of the magnetic moment $\mu_z$ carried by electrically generated droplet soliton. Voltage pulse with the height $V = 4$~V and duration $\tau_V = 0.3$~ns is applied to the gate electrode with the radius $R_G = 50$~nm, and the magnetic field strength equals 300~Oe. The inset shows the spatial distribution of the magnetization direction cosine $m_z(x, y)$ under the gate electrode at $t = 0.28$~ns. }
\end{figure}

In the study of electrically induced magnetization dynamics, we consider the MgO/Fe/MgO heterostructures subjected to rectangular voltage pulses with the duration $\tau_V$ ranging from 0.1 to 1 ns. The pulse amplitude is set to -4~V, at which the electric field $E_z$ in the 2-nm-thick MgO nanolayer is below its breakdown field $E_b \approx 2.4$~V~nm$^{-1}$~\cite{Dimitrov:2009}.The micromagnetic simulations demonstrate that a voltage pulse providing local reduction of PMA may induce precessional magnetization switching by about 180$^\circ$ inside the Fe region beneath the gate electrode. The switched region occupies an area $S$ smaller than the gate area $S_G$ and may have nearly a disk shape (see the inset in Fig.~\ref{fig:mu-t-300Oe-50nm}). The switching creates a change $\mu$ in the magnetic moment of the Fe film, which has a dominant out-of-plane component $\mu_z(t) =- M_s t_\mathrm{F} \displaystyle\int_{S_G} dx dy [m_z(\mathbf{r}, t) - m_z(\mathbf{r}, t=0)]$. Since $\mu_z(t = \tau_V) \sim 2 M_s t_\mathrm{F} S$, the switched region represents a magnetic droplet soliton. Figure~\ref{fig:mu-t-300Oe-50nm} shows a representative time dependence of the droplet moment $\mu_z(t)$. We see that $\mu_z$ oscillates with a gradually increasing frequency $f_\mu(t)$ and decreasing amplitude and becomes negligible after a few nanoseconds. The analysis of the simulation data reveals that such a behavior is mostly due to the oscillations of the droplet area $S$. Importantly, the soliton does not experience any significant drift towards the boundary of the Fe disk during the whole period of its existence. 

\begin{figure}[!h]
\centering
\includegraphics[height=8cm]{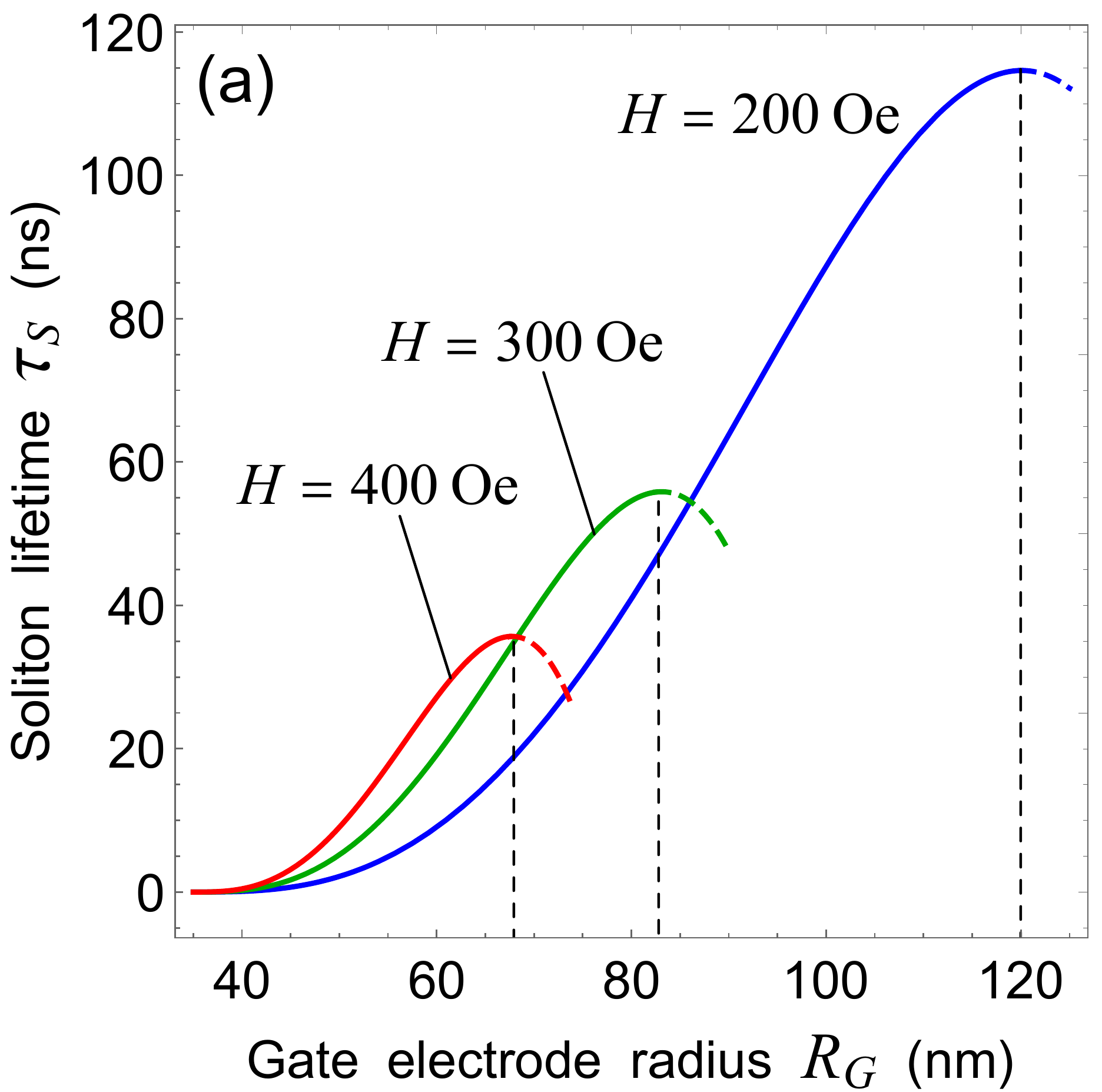}
\includegraphics[height=8cm]{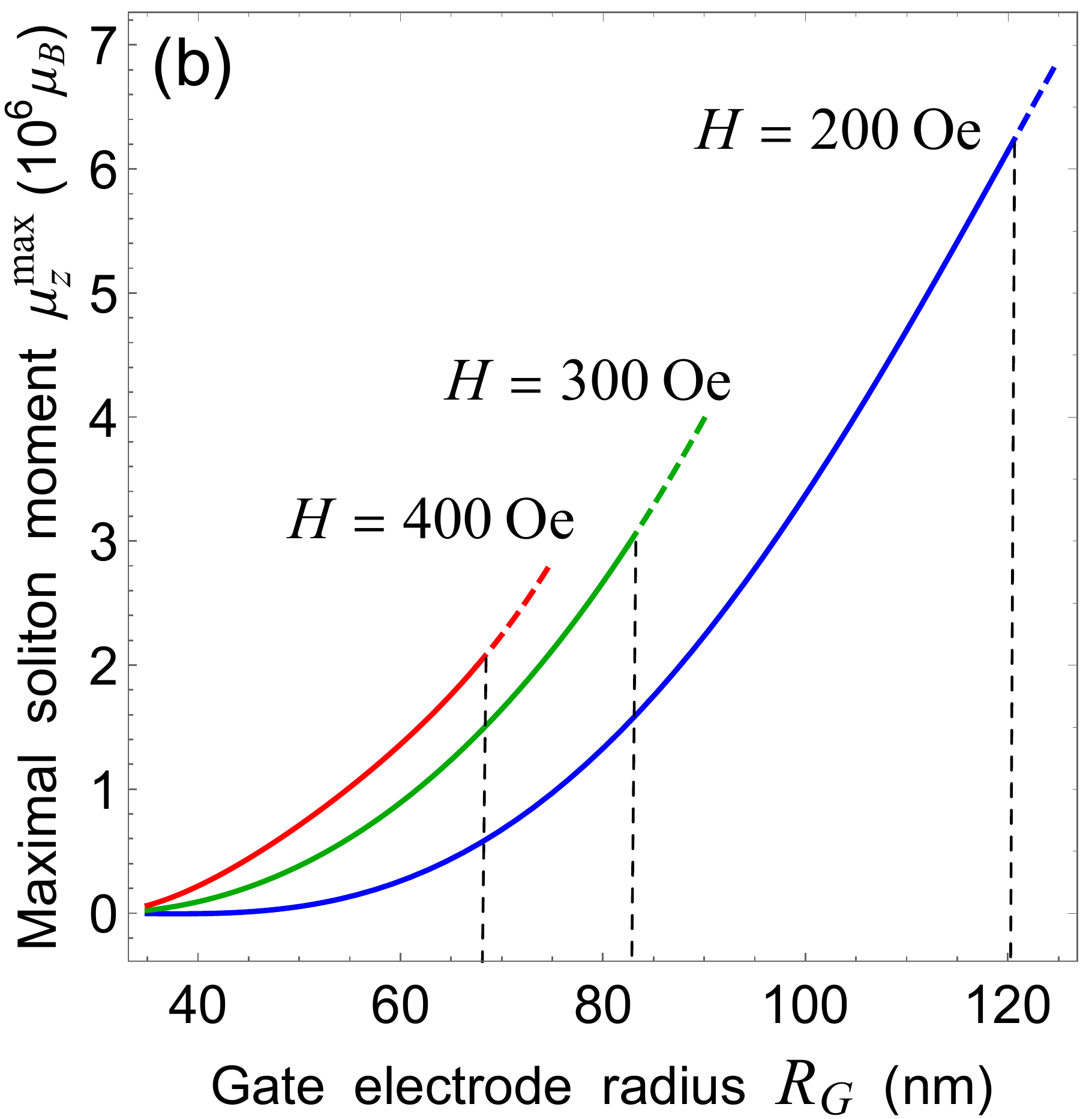}
\caption{\label{fig:t-Rg} Soliton lifetime $\tau_S$ (a) and maximal value of the soliton magnetic moment $\mu_z$ (b) plotted as a function of the gate electrode radius $R_G$. The strengths $H$ of the applied magnetic field are indicated on the plots.  }
\end{figure}

The soliton lifetime $\tau_S$ depends on the pulse duration $\tau_V$, gate radius $R_G$, and strength of external magnetic field $\mathbf{H}$. By analyzing the results of simulations performed at different values of $\tau_V$, we find that the optimal pulse duration $\tau_V^*$, which maximizes the soliton lifetime, corresponds to the minimum of the quantity $\displaystyle \int_{S_G} dx dy \bigg(H_z^\mathrm{eff} - \displaystyle \frac{2 k_s V}{\mu_0 M_s t_\mathrm{MgO} t_\mathrm{F}}m_z\bigg)$. At a fixed magnetic field, the optimal duration $\tau_V^*$ first grows with the increasing gate radius $R_G$ (see Appendix). However, above some threshold radius $R_\mathrm{th}$ two or three solitons form under the gate instead of one. In what follows we present only the results of simulations performed for the heterostructures involving gates with $R_G < R_\mathrm{th}$ subjected to voltage pulses of the optimal duration $\tau_V^*(R_G, H)$.

Variations of the soliton lifetime $\tau_S$ with the gate radius and the field strength are presented in Fig.~\ref{fig:t-Rg}(a). 
At a fixed field strength, $\tau_S$ grows with increasing gate size up to some optimal radius $R_G^*(H)$, at which $\tau_S$ reaches maximal value. Remarkably, the soliton lifetime attained at $H = 200$~Oe exceeds 100~ns when the gate radius is close to  $R_G^*(H = 200 \; \text{Oe}) \approx 120$~nm. As may be expected, the growth of $\tau_S$ at $R_G < R_G^*$ correlates with the dependence of the maximal value of the soliton moment $\mu_z(t)$ on the gate radius [see Fig.~\ref{fig:t-Rg}(b)]. The decrease of $\tau_S$ at $R_G > R_G^*$ may be attributed to the enhancement of spin-wave radiation by larger droplets, which takes the magnetic moment $\mu_z(t)$ away from the soliton (see Appendix). 

To evaluate the influence of thermal fluctuations on the soliton lifetime, we carry out additional simulations with the account of a stochastic Gaussian noise. In these simulations, the effective field $\mathbf{H}_\mathrm{eff}$ involved in the LLG equation includes a thermal random field $\mathbf{H}_\mathrm{th}$ in the form employed by the MuMax3 software~\cite{mumax3}. The results show that the introduction of $\mathbf{H}_\mathrm{th}$ corresponding to 300~K reduces $\tau_S$ approximately by a factor of two. Hence the soliton can live up to 50~ns at the room temperature.

\subsection{Propagation and routing of droplet solitons}
\begin{figure}[!h]
\centering
    \includegraphics[height=3cm]{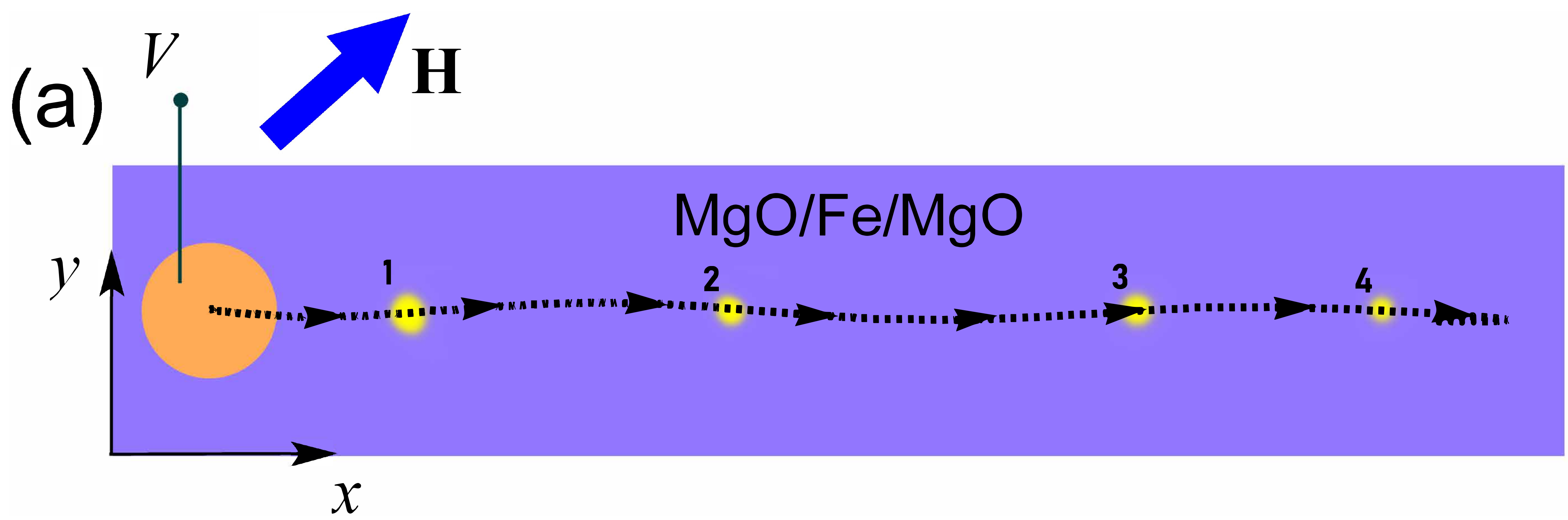}
    \vfill
    \includegraphics[height=8cm]{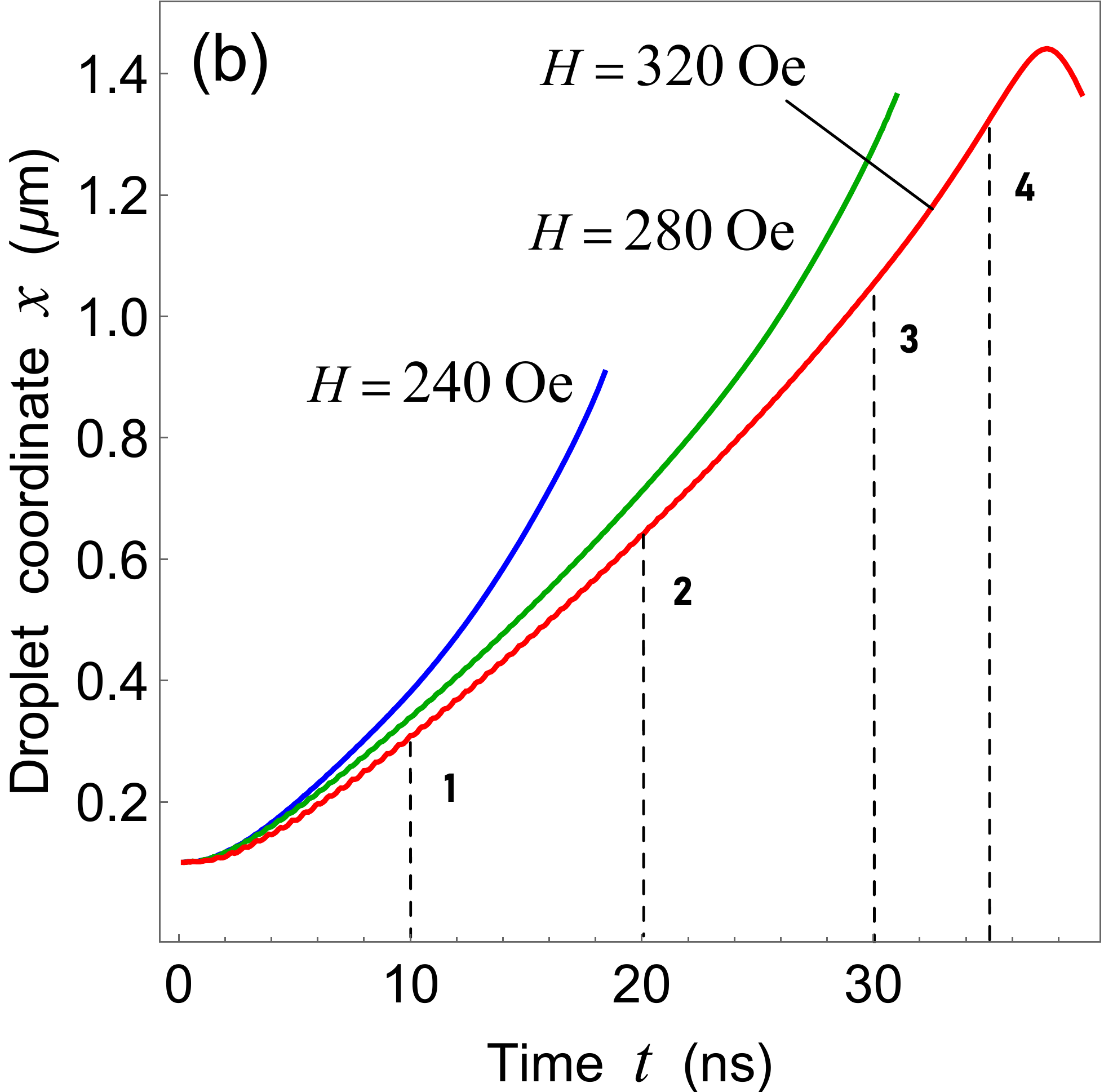}
    \caption{\label{fig:propagation} Propagation of electrically generated droplet soliton in the MgO/Fe/MgO waveguide. Panel (a) shows a representative trajectory of the soliton formed at $H = 320$~Oe under circular gate electrode placed near the waveguide edge. Yellow regions depict the droplet area at four different positions. Panel (b) presents the droplet position $x$ in the Fe microstripe as a function of time at different field strengths $H$ indicated on the plot. The radius $R_G$ of gate electrode equals 70~nm. The curves break when the droplet magnetic moment goes to zero. }
\end{figure}
Next, it is important to determine how far the generated droplet soliton can propagate along a ferromagnetic waveguide. To address this question, we carry out micromagnetic simulations for the MgO/Fe/MgO trilayer with a rectangular shape, representative IP dimensions $L_x = 1.5$~$\mu$m and $L_y = 300$~nm, and a circular gate electrode placed near the waveguide beginning [see Fig.~\ref{fig:propagation}(a)]. In such a heterostructure, the generated soliton moves away from the nucleation region beneath the gate electrode owing to the existence of a gradient $\partial \mathbf{H}_\mathrm{dip} / \partial x$ of the demagnetizing field $\mathbf{H}_\mathrm{dip}$~\cite{Hoefer2012}, which is nonuniform in the rectangular Fe microstripe. As a result, the soliton propagates along the waveguide, experiencing small deviations from its central line [Fig.~\ref{fig:propagation} (a)], which are caused by initial misalignment of the droplet velocity and restoring forces created by the microstripe edges.

Figure~\ref{fig:propagation} (b) shows the droplet position in the waveguide as a function of time at different external magnetic fields. It is seen that the propagation distance grows with increasing field strength $H$, exceeding one micrometer at $H > 250$~Oe. This behavior is caused by the field-induced increase of the soliton lifetime in the Fe microstripe, which overcompensates the decrease in the average soliton velocity ranging from 49.3~m~s$^{-1}$ at $H = 240$~Oe to 38.5~m~s$^{-1}$ at $H = 320$~Oe. As the departure droplet size grows with increasing field strength, we arrive at the conclusion that small solitons move faster than large ones. 

\begin{figure}[!h]
    \center{
    \includegraphics[width=0.85\linewidth]{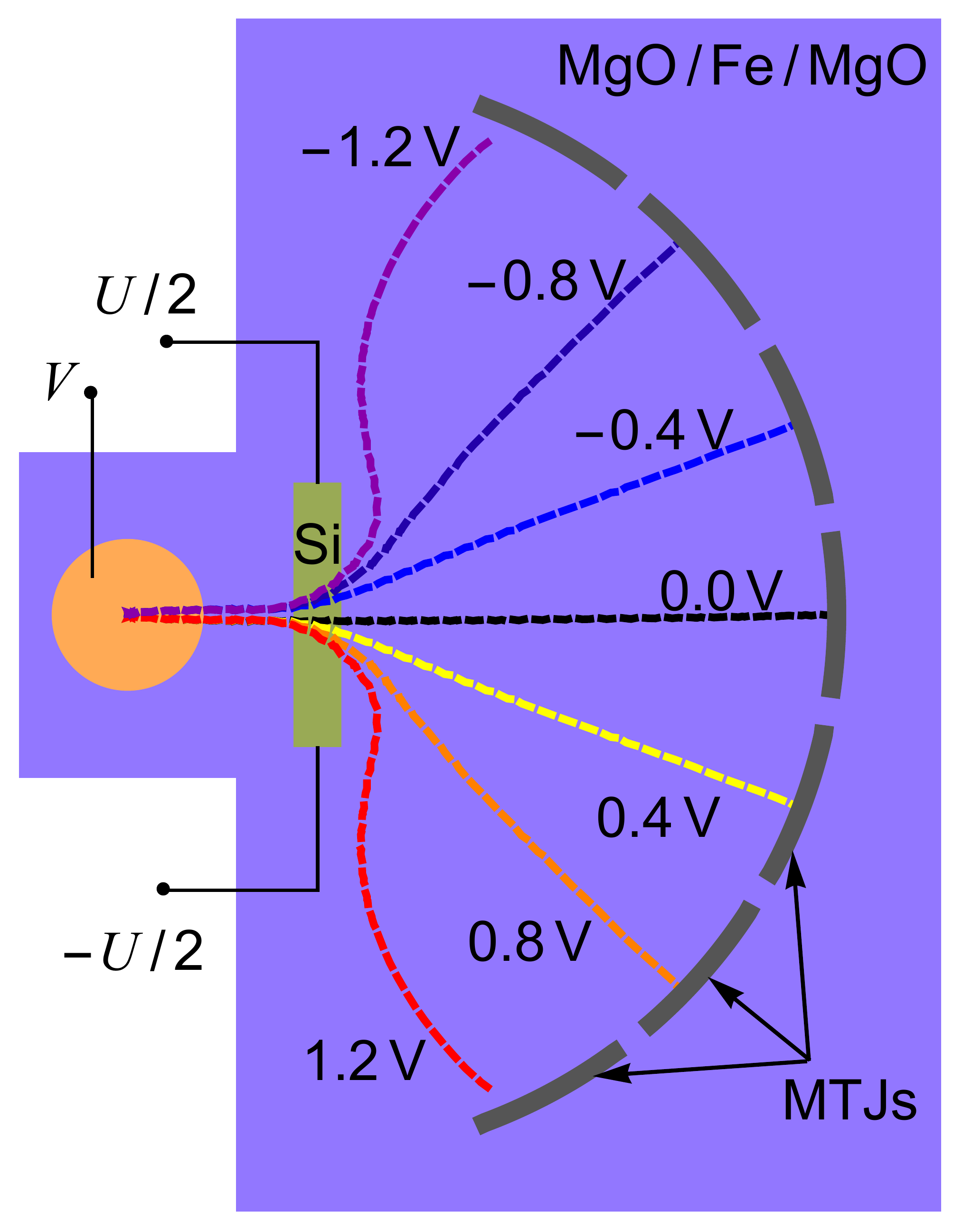}}
    \caption{\label{fig:routing-traject} Routing of droplet solitons in the MgO/Fe/MgO-based structure comprising circular gate electrode and semiconducting nanostripe electrode placed on the upper MgO nanolayer. Dashed lines show the trajectories of the droplet center predicted by micromagnetic simulations at $H = 300$~Oe. Magnitude $U$ of dc voltages $U/2$ and $-U/2$ applied to the ends of semiconducting nanostripe is indicated near the corresponding path line. Gray arcs depict ferromagnetic electrodes deposited on the upper MgO nanolayer at the distance $D = 0.5$~$\mu$m from the nanostripe center. Together with the extended Fe interlayer, these electrodes form magnetic tunnel junctions enabling electrical detection of the soliton trajectory. }
\end{figure}

Finally, we describe an efficient method of electrical routing of the droplet solitons to different outputs of a spintronic device. Such a routing can be realized in the MgO/Fe/MgO-based structure shown in Fig.~\ref{fig:routing-traject}, where additional semiconducting nanostripe electrode is placed on the upper MgO nanolayer near the circular gate electrode. The application of dc voltages $U/2$ and $-U/2$ to the ends of semiconducting nanostripe gives rise to an electric current flowing along the electrode, which creates a linear variation of the voltage applied to the underlying MgO area.
\noindent
As a result, a voltage-controlled gradient of PMA appears in the Fe region beneath the semiconducting electrode. The simulations reveal that the PMA gradient strongly affects the soliton trajectory in the Fe film owing to additional acceleration of the droplet in the direction antiparallel to the gradient vector. Figure~\ref{fig:routing-traject} shows the trajectories of the soliton generated by the gate electrode with the radius $R_G = 70$~nm and routed by the Si nanostripe with the length $\delta_y = 200$~nm, width $\delta_x = 50$~nm, and resistivity $\rho = 83 $~m$\Omega$~m~\cite{Si-rho}. When voltages $U/2$ and $-U/2$ are applied to the nanostripe ends, the trajectory of the droplet center changes beneath the nanostripe, deviating from approximately straight path forming at $U = 0$. The direction of this deviation depends on the sign of $U$, and the deviation magnitude can be characterized by an angle $\beta$ between the straight soliton trajectory and a line connecting the nanostripe center and the droplet position at a fixed distance $D$ from this point. 
\begin{figure}[!h]
    \center{
    \includegraphics[width=0.95\linewidth]{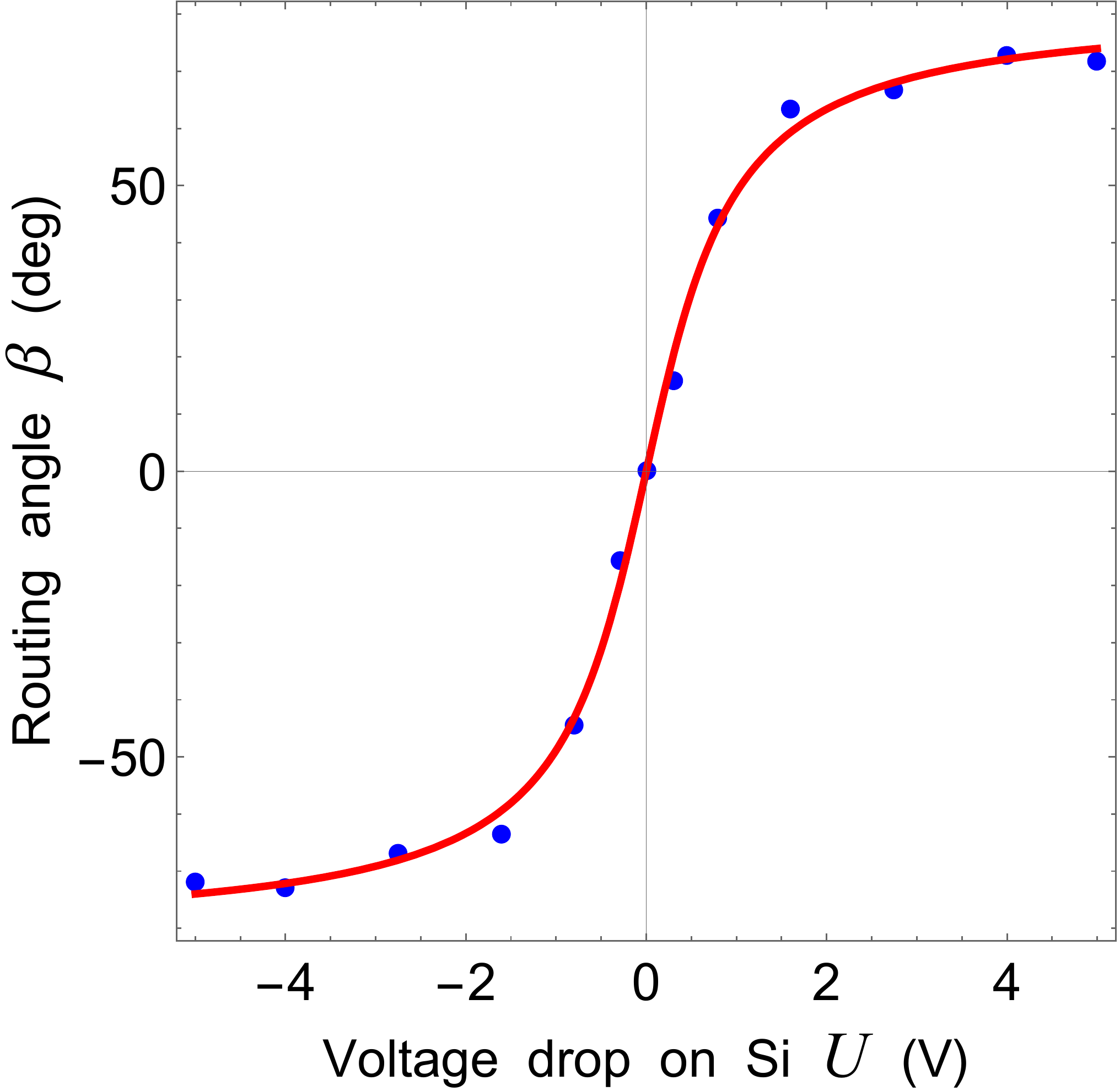}}
    \caption{\label{fig:alpha-U} Dependence of the soliton routing angle $\beta$ on the magnitude $U$ of dc voltages $U/2$ and $-U/2$ applied to the ends of Si nanostripe with the length $\delta_y = 200$~nm and width $\delta_x = 50$~nm. Simulation data (points) are fitted by the function $\beta = 52^\circ \; \mathrm{arctan}(1.4 U / \text{V})$ (curve). }
\end{figure}
\noindent
Figure~\ref{fig:alpha-U} presents the voltage dependence of the routing angle $\beta(U)$ determined at $D = 0.5$~$\mu$m. Remarkably, the routing angle reaches about 50$^\circ$ at $U = 1$~V, varying almost linearly up to $U = 0.8$~V with the mean rate $d \beta / d U \approx 56^\circ$~V$^{-1}$. At voltages $U > 1$~V, the dependence $\beta(U)$ becomes nonlinear. This feature may be attributed to the arising significant influence of the sample edges on the soliton path, which is evidenced by the curved droplet trajectories obtained at $U = \pm 1.2$~V (see Fig. ~\ref{fig:routing-traject}).

The demonstrated electrical control of the soliton trajectory makes it possible to transfer the magnetic signal to one of several outputs of the device. The signal can be read electrically with the aid of a magnetic tunnel junction (MTJ) formed by a nanoscale ferromagnetic electrode deposited on the upper MgO nanolayer and the underlying region of the Fe interlayer. Indeed, owing to the phenomenon of spin-dependent tunneling, the MTJ resistance changes strongly after the magnetization reversal in one of ferromagnetic electrodes~\cite{TMR}. Therefore, the soliton appearance in the Fe region below the perpendicularly magnetized top ferromagnetic electrode will manifest itself in a resistance change, which can be easily detected electrically.

\section{Conclusions}
In summary, we theoretically studied the electrically driven magnetization dynamics in the MgO/Fe/MgO trilayer with the voltage-controlled magnetic anisotropy. The micromagnetic simulations demonstrated that the application of a sub-nanosecond voltage pulse to the nanoscale gate electrode placed on the MgO nanolayer gives rise to the formation of the magnetic droplet soliton despite the presence of nonzero magnetic damping. The soliton lifetime, which depends on the gate size and the strength of in-plane external magnetic field, can reach 50~ns at room temperature and 100~ns in the absence of thermal fluctuations. When generated near the edge of the Fe microstripe, the soliton can propagate over a distance exceeding one micrometer with the mean speed about 40~m~s$^{-1}$ owing to the existing gradient of the demagnetizing field. By passing a small electric current density $\sim 10^8$~A~m$^{-2}$ along additional Si nanostripe electrode, we also achieved an efficient electrical routing of the soliton in the extended Fe interlayer.

Our theoretical results provide guidelines for the development of an energy-efficient information-processing device based on the electrical generation, propagation, and routing of magnetic solitons. The device converts the input voltage signal into the magnetic information carrier, which propagates to one of several outputs. The desired output is selected by the voltage applied to the routing electrode and involves the magnetic tunnel junction, which provides electrical reading of the output signal via the measurement of the junction's resistance. 

\bibliography{ref.bib}

\newpage
\appendix*
\section{}
\renewcommand\thefigure{\thesection ~A\arabic{figure}} 
\setcounter{figure}{0}   

\begin{figure}[h]
    \center{
    \includegraphics[width=0.8\linewidth]{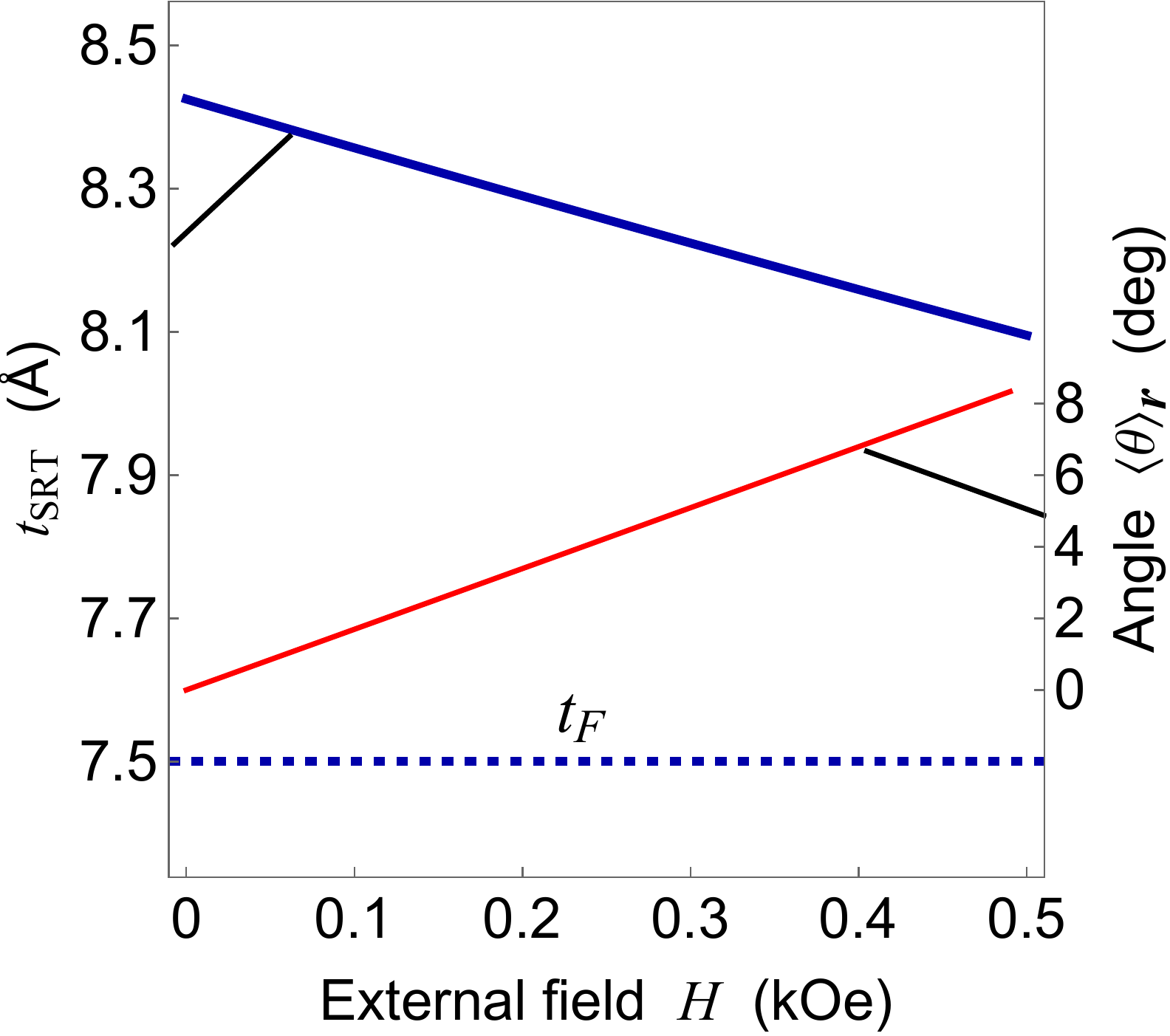}}
    \caption{\label{fig:tSRT} The critical Fe thickness $t_\mathrm{SRT}$ and the mean value $\langle \theta \rangle$ of the magnetization polar angle $\theta$ at $t_\mathrm{F} = 0.75$~nm (dashed line) plotted as a function of the magnetic field strength $H$. }
\end{figure}

To confirm that the Fe film in the MgO/Fe/MgO trilayer retains almost perpendicular-to-plane orientation in all performed micromagnetic simulations, we calculated the critical Fe thickness $t_\mathrm{SRT}$ and the mean value $\langle \theta \rangle$ of the magnetization polar angle $\theta$ as a function of the strength $H$ of applied in-plane magnetic field. The results presented in Fig.~\ref{fig:tSRT} show that $t_\mathrm{SRT}$ exceeds the Fe thickness $t_\mathrm{F} = 0.75$~nm and the mean value $\langle \theta \rangle$ is smaller than 8$^\circ$ even at the highest field $H = 400$~Oe used in the simulations.
\begin{figure}[h]
    \center{
    \includegraphics[width=0.8\linewidth]{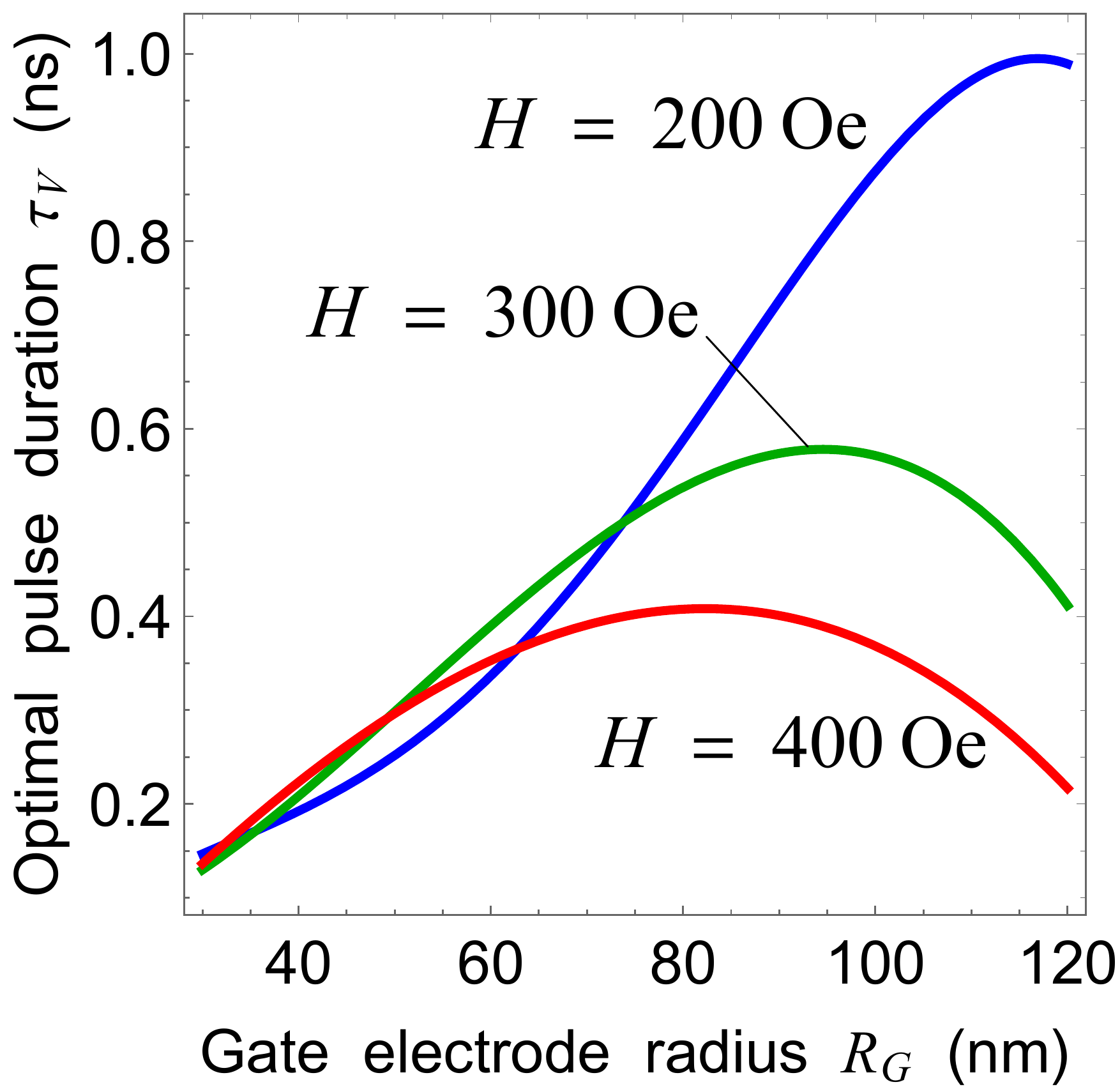}}
    \caption{\label{fig:tau:RG} Dependences of optimal pulse duration $\tau_V^*$ on gate radius $R_G$ calculated at different strengths $H$ of applied magnetic field. }
\end{figure}

The optimal duration $\tau_V^*$ of the voltage pulse, which maximizes the soliton lifetime $\tau_S$, depends on the size of the gate electrode and on the magnetic field strength. Figure~\ref{fig:tau:RG} demonstrates variations of $\tau_V^*$ with the gate radius $R_G$ calculated at three different field strengths. It is seen that the optimal pulse duration first increases with the growing gate radius but begins to decrease when $R_G$ exceeds some threshold value. Such a change in the dependence $\tau_V^*(R_G)$ may be attributed to the appearance of inhomogeneous magnetization switching under larger gate electrodes.

\begin{figure*}[httb]
\centering
\includegraphics[width=1\linewidth]{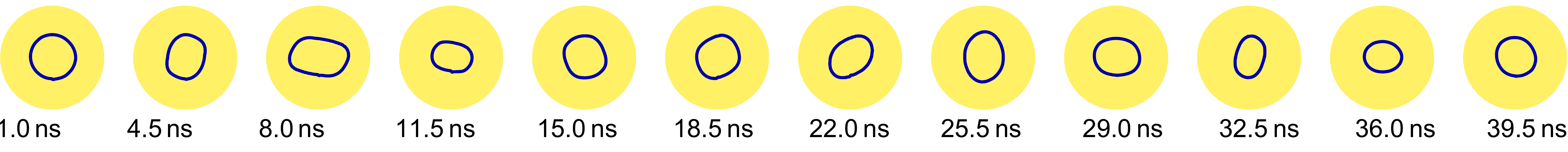}
\vfil
\includegraphics[width=1\linewidth]{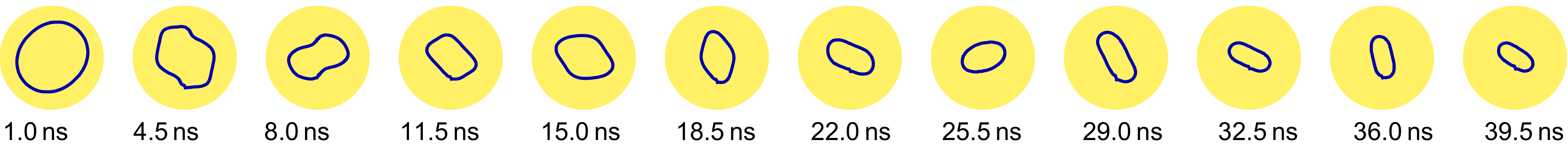}
\caption{\label{fig:boundaries} Temporal evolutions of droplet solitons generated by optimal voltage pulses applied to gate electrodes with radius $R_G = 90$~nm (upper panel) and $R_G =110$~nm (lower panel). The black line shows the soliton boundary, while the yellow area indicates the Fe region beneath the gate electrode. The magnetic field strength $H = 200$~Oe. }
\end{figure*}

The simulations also show that the gate radius $R_G$ affects both the soliton size and shape. Figure~\ref{fig:boundaries} compares the temporal evolutions of the droplets formed under the gate electrodes with different diameters. It can be seen that both solitons experience significant shape variations during the decay process. However, the droplet generated by the voltage pulse applied to the electrode with the smaller radius $R_G = 90$~nm becomes nearly circular when the time approaches 40~ns. In contrast, the soliton formed under the larger gate with $R_G = 110$~nm has a strongly anisotropic shape and fluctuating boundary at this time. This feature gives rise to a high-power spin-wave radiation from the soliton, which causes rapid reduction of its area and magnetic moment.


\end{document}